\documentclass[aps,preprint]{revtex4}%
\usepackage{amsfonts}
\usepackage{amsmath}
\usepackage{amssymb}
\usepackage{graphicx}%
\begin{document}
\title{High energy head-on particle collisions near event horizons: classification of scenarios}
\author{H.V. Ovcharenko}
\affiliation{Department of Physics, V.N.Karazin Kharkov National University, 61022 Kharkov, Ukraine\\}
\affiliation{Institute of Theoretical Physics, Faculty of Mathematics and Physics, Charles
University, Prague, V Holesovickach 2, 180 00 Praha 8, Czech Republic}
\author{O.B. Zaslavskii}
\affiliation{Department of Physics and Technology, Kharkov V.N. Karazin National
University, 4 Svoboda Square, Kharkov 61022, Ukraine}

\begin{abstract}
We consider head-on collisions of two particles near the event horizon.
Particle 1 is outgoing, particle 2 is ingoing. We elucidate the conditions
when the energy $E_{c.m.}$ in the center of mass frame can grow unbounded. In
doing so, we dispel some misconceptions popular in literature on high energy
collisions near black holes and carry out clear distinction when such a
collision happens near a black hole and when this occurs near a white one. If
the proper time between the horizon and an arbitrary point outside it for
particle 1 is finite, we deal with a white hole. If it is infinite, we deal
with a black hole. Particles can be either free or experience the action of a
finite force. Our results are complementary to those for the standard BSW
effect when particles move in the same direction. The results rely on
classification of particles developed in our previous work H.V. Ovcharenko,
O.B. Zaslavskii, Phys. Rev. D 108, 064029 (2023). The aforementioned article
performed the first step towards a building classification of all possible
high energy collisions near black/white hole for different types of horizons
and nonzero force. In the present one, we perform the second step.

\end{abstract}
\maketitle
\tableofcontents

\section{Introduction}

During last 15 years a new direction in black hole physics has been developing
actively. This happened after finding of Ba\~{n}ados, Silk and West called the
BSW effect \cite{ban}. According to \ it, if two particles move towards a
black hole and collide near the black hole horizon, the energy $E_{c.m.}$ in
the center of mass frame can grow unbounded if parameters of one particle
(called critical) are fine-tuned.

Publication of the aforementioned paper drew attention to the fact that the
high energy processes near the horizon were already considered in the series
of papers before \cite{pir1} - \cite{pir3} for the Kerr metric. However, in
these pioneering works \cite{pir1} - \cite{pir3} with interesting and
important results, there were no explicite indication of the nature of
background for which corresponding findings are valid. In particular, in eqs.
2.54 of \cite{pir3} the divergence of energy $E_{c.m.}$ for head-on collision
was found. Meanwhile, if in the background with a nonextremal horizon a
particle is outgoing (moves in the outward direction), it emerges from a
white, not black hole (see in more detail below). Failure to distinguish both
cases has led to confusion between black and white holes and provoked a series
of misleading statements in consequent literature in \ which the BSW effect
and the results of \cite{pir1} - \cite{pir3} were considered on the same
footing. And this is despite the fact that energetics of both processes are
completely different. In particular, the BSW effect requires fine-tuning
\cite{ban}, \cite{prd}, whereas this requirement is absent for head-on
collisions. From another hand, the results of \cite{pir1} - \cite{pir3}, if
interpreted properly, can have potential important consequences concerning the
instability of white holes. 

It is worth stressing that conclusion about the white (not black) hole nature
of the corresponding horizon in the \ context under discussion applies to
nonextremal metrics. If it is extremal, outgoing trajectories are possible in
a black (not white) \ hole background. Moreover, they lead to essential
enhancement of energy extraction as compared with what happens in the original
BSW\ scenario in the Kerr background \cite{sch}.

Findings \cite{ban} - \cite{pir3} created hope to relate the corresponding
theoretical predictions with realistic processes in astrophysics including
gamma-ray burst, jets, etc. and elucidate potential connections with dark
matter particles. Up to date, realiable observational confirmation of
aforementioned mechanism is still absent. However, the effect of high energy
collisions near horizon is interesting theoretical result that has universal
character and is valid not only for massive astrophysical black holes. It
looks necessary to build a theory of this effect as complete as possible that
would facilitate further interpretation of observations. In turn, this
requires careful analysis of all possible scenarios of collisions including
clear distinction between black and white holes.

Our goal is to generalize the results for ultra-high energy collisions for
metrics with a horizon, classify them and describe in which cases these
process occur near a black hole, and in which ones they concern white holes.
We do not discuss specially the question of instability of white hole (see
Sec. 15.2 in \cite{fn} and references therein) that is a separate subject
beyond the scope of our paper, but we develop formalism that can be used in
such an analysis. It is worth noting that general approach to classification
of scenarios of high energy collisions (but for the Kerr metric only) was
suggested recently in \cite{del2} but also without clear distinction between
white and black holes.

Originally, motivation for studying high energy collisions came from
astrophysics where they were discussed, for example, as potential source as
gamma-ray bursts \cite{pir2}. Unfortunately, until now, there is no realistic
confirmation of this astrophysical realization. Doubts become even stronger if
one takes into account necessity of white holes for some scenarios.

We deem that, first of all, full classification of all possible scenarios is
necessary that can be, in principle, used in astrophysical applications
further. Meanwhile, even without such applications, high energy collisions
near black/white holes is a nontrivial physical effect that should be studied
as completely as possible. In general, there are three main cases for
classification: (i) both particles move towards the horizon (BSW effect), (ii)
they move in opposite directions (head-on collision), (iii) one of particles
moves along the near-horizon circle orbit. Case (i) was considered in our
previous work \cite{force23}, case (ii) is the subject of the current
manuscript. We hope to consider case (iii) elsewhere. Additionally, we note
that our consideration below (as well as in (i)) remains valid even if a force
acts on a particle.

\section{General setup}

\subsection{Metrics}

We are investigating the motion of particles in the background of a rotating
black hole which is described in the generalized Boyer-Lindquist coordinates
$(t,r,\theta,\varphi)$ by the metric:%
\begin{equation}
ds^{2}=-N^{2}dt^{2}+g_{\varphi\varphi}(d\varphi-\omega dt)^{2}+\frac{dr^{2}%
}{A}+g_{\theta\theta}d\theta^{2},
\end{equation}
where all metric coefficients do not depend on $t$ and $\varphi$. The horizon
is located at $r=r_{h}$ where $A(r_{h})=N(r_{h})=0$. Near the horizon, we use
a general expansion for the functions $N^{2}$, $A$ and $\omega$:%
\begin{equation}
N^{2}=\kappa_{p}v^{p}+o(v^{p}),\text{ \ \ }A=A_{q}v^{q}+o(v^{q}),
\label{an_exp}%
\end{equation}%
\begin{equation}
\omega=\omega_{H}+\omega_{k}v^{k}+o(v^{k}), \label{om_exp}%
\end{equation}
where $q,p$ and $k$ are integers that characterize the rate of a change of the
metric functions near the horizon, and $v=r-r_{h}.$

By definition, if $p=q=1$, a horizon is called nonextremal, if $q=2$, $p\geq
2$, the horizon is extremal. If $q\geq3$, $p\geq2$, the horizon is
ultraextremal. The requirement of regularity imposes some restrictions on
these numbers (see \cite{reg}, especially Tables 1 and 2 there).

\subsection{Motion of free particles}

Now, let us investigate the motion of a particle in such a space-time. If a
particle is freely moving, the space-time symmetries with respect to
$\partial_{t}$ and $\partial_{\varphi}$ impose conservation of the
corresponding components of the four-momentum: $mu_{t}=-E$, $mu_{\varphi}=L$.
We assume the symmetry with respect to the equatorial plane. In what follows,
we restrict ourselves by equatorial motion. Then,%
\begin{equation}
u^{t}=\frac{\mathcal{X}}{N^{2}}\text{,} \label{t}%
\end{equation}%
\begin{equation}
u^{\varphi}=\frac{\mathcal{L}}{g_{\varphi\varphi}}+\frac{\omega\mathcal{X}%
}{N^{2}}\text{.} \label{phi}%
\end{equation}
Using also the normalization condition $u_{\mu}u^{\mu}=-1$, we obtain that%
\begin{equation}
u^{r}=\sigma\frac{\sqrt{A}}{N}P, \label{r}%
\end{equation}
where $\sigma=\pm1$,
\begin{equation}
\mathcal{X}=\epsilon-\omega\mathcal{L}\text{,} \label{hi}%
\end{equation}
$\epsilon=E/m$, $\mathcal{L}=L/m$ and $P$ is given by:%
\begin{equation}
P=\sqrt{\mathcal{X}^{2}-N^{2}\left(  1+\frac{\mathcal{L}^{2}}{g_{\varphi
\varphi}}\right)  }. \label{P}%
\end{equation}

This allows us to write the four-velocity of a free-falling particle in the
following form:%
\begin{equation}
u^{\mu}=\left(  \frac{\mathcal{X}}{N^{2}},\sigma\frac{\sqrt{A}}{N}%
P,0,\frac{\mathcal{L}}{g_{\varphi\varphi}}+\frac{\omega\mathcal{X}}{N^{2}%
}\right)  . \label{4_vel}%
\end{equation}

\subsection{Motion of accelerated particles}

Now, let a particle move non-freely. In the case of an external force acting
on the particles, the quantities $\varepsilon$ and $\mathcal{L}$ are obviously
not conserved. However, despite this fact, we can still use the expression
(\ref{4_vel}) but with general functions $\mathcal{X(}r\mathcal{)}$ and
$\mathcal{L(}r\mathcal{)}$, provided forces do not depend on time and angle
variables. This can be substantiated as follows.

By definition, we introduce the energy and angular momentum according to
$E=-mu_{t}$, $L=mu_{\varphi}$. Then, using $u^{t}=g^{t\mu}u_{\mu}$ we obtain
eq. (\ref{t}) immediately. In a similar way, for $u^{\varphi}=g^{\varphi\mu
}u_{\varphi}$ we obtain (\ref{phi}). And, from the normalization condition
$u_{\mu}u^{\mu}=-1$we get (\ref{r}).

Near the horizon, we can use the Taylor expansion for them:%
\begin{equation}
\mathcal{X}=X_{0}+X_{s}v^{s}+o(v^{s}),\text{ \ \ }\mathcal{L=}L_{H}+L_{b}%
v^{b}+o(v^{b}). \label{L_exp}%
\end{equation}

If a particle is fine-tuned, $X_{0}=0$ by definition.

Detailed analysis of near-horizon particle dynamics and the BSW effect under
the action of a force was given in \cite{tz13} for extremal black holes and in
\cite{tz14} for nonextremal ones. More general analysis of the BSW effect for
different types of horizons with a force was given in \cite{force23}.

\section{Kinematics of particle motion near the horizon}

In this section, we discuss how the properties of particle motion are
connected with the type of horizon and elucidate the key role of the proper time.

\subsection{Nonextremal horizon}

To begin with, let us consider a nonextremal horizon. Is it possible for a
usual (not-fine-tuned) particle having a finite energy to move in the outward
direction if this motion starts (or continues) in the immediate vicinity of a
horizon of a black hole? On the first glance, nothing prevents it. However,
more close inspection shows that this cannot be realized.

If a particle has finite $E$, this means that the proper time between the
horizon and its point on a trajectory outside it is finite. In turn, it means
that if we continue a trajectory in the past, it crossed the horizon. Thus
this particle appeared from the inner region under the horizon. But this is
nothing else than a white (not black) hole horizon in contradiction with assumption.

Naively, one could try to avoid this difficulty and suggest, say, such a
scenario. Some particle comes from inifnity, bounces back near the horizon and
returns to inifnity. However, if there is turning point (in the sense of
\ radial motion) on trajectory, it follows that in this point $P=0$. Then,
according to (\ref{P}), $X\sim N$ in this point. For an intermediatepoint with
$N=O(1)$ this is possible. However, if we take the horizon limit
$N\rightarrow0$, this entails $X\rightarrow0$ as well. But \ this is
inconsistent with the condition that a particle is usual. Thus we must reject
this scenario.

Another attempt may consist in taking the near-critical particle for which
$X=0$ on the horizon. However, such a particle cannot reach the nonextremal
horizon since $X=O(v)=O(N^{2})$, so the expression inside the square root
becomes negative. Instead of the exactly critical particle one can take a
near-critical one but the correspondinbg area in which motion is possible
shrinks in this limit \cite{gp}, \cite{prd}, \cite{near}.

There is one more option. If one send a supermassive particle from infinity,
it can create a new usual (not fine-tuned) outgoing but the price for this is
too high. Namely, one of initial particles should have very large mass and
energy, $E\gtrsim m\sim N_{c}^{-2}$(subscript "c" refers to the point of
collision). As input itself is tremendous, this invalidates the scenario
completely. These results were obtained in \cite{com} for the Kerr metric and
generalized in \cite{epl1}. Moreover, if all energies and masses are finite,
it is proven in \cite{epl1} that near-horizon collision of any number of
ingoing particles  cannot produce a usual outgoing one. In other words, if in
the initial configuration usual outgoing particles are absent, they cannot
appear after collision.

The main conclusion that follows from the above discussion consists in that a
usual particle of a finite mass emerges from the \textit{white} hole region.
The proper time for such a particle is finite.

\subsection{Behavior of proper time for different types of particles and
horizon\label{sec_pr_time}}

In this subsection, we generalize the material of the previous ine and
describe briefly, which particles participate in collision. First of all, we
assume that in this process fine-tuned particles may participate. They are
classified according to the approach developed in \cite{force23}. Below, we
take advantage of corresonding results given in Sec. IV and, especially, Table
I of \ the aforementioned paper.

Particle is called usual if expansion for $\mathcal{X}$ (\ref{L_exp}) starts
with the nonzero constant term $X_{0}.$

Particle is called subcritical if for such a particle expansion (\ref{L_exp})
holds with $0<s<p/2.$ Particle is called critical if for such a particle
expansion (\ref{L_exp}) holds with $s=p/2.$ Particle is called ultracritical
if for such a particle (\ref{L_exp}) holds with $s=p/2$ and additionally
coefficients $X_{s}$ satisfy condition:
\begin{equation}
(X_{s}v^{s})^{2}-\kappa_{p}\left(  1+\frac{L_{H}^{2}}{g_{\varphi H}}\right)
v^{p}\approx\frac{\kappa_{p}}{A_{p}}(u^{r})_{c}^{2}v^{2c+p-q}\text{,}
\label{uc}%
\end{equation}

where $c>q/2.$ Here, we put $u\approx(u^{r})_{c}v^{c}$ where $(u^{r})_{c}$ are
some constants. It follows from (\ref{uc}) with $s=p/2$ that%
\begin{equation}
(X_{s})^{2}-\kappa_{p}\left(  1+\frac{L_{H}^{2}}{g_{\varphi H}}\right)
\approx\frac{\kappa_{p}}{A_{p}}(u^{r})_{c}^{2}v^{2c-q}\text{.}%
\end{equation}

In addition to the behavior of the four-velocity we are going to discuss the
behavior of a proper time. Proper time for motion from point $r=r_{1}$ to
$r=r_{2}$ is given by:%

\begin{equation}
\tau=\int_{r_{2}}^{r_{1}}\frac{dv}{u^{r}},
\end{equation}
if $r_{2}<r_{1}$. If $r_{2}>r_{1}$, the lower and upper limits in integral interchange.

\bigskip For usual particles, using (\ref{P}) and (\ref{4_vel}), one obtains
when $r_{2}\rightarrow r_{h}$:%
\begin{equation}
\tau\sim\int\frac{dv}{v^{\frac{q-p}{2}}}\sim v^{\frac{p-q}{2}+1}.
\end{equation}

The proper time is finite if $q<p+2.$

For subcritical particles one obtains%
\begin{equation}
\tau\sim\int\frac{dv}{v^{s+\frac{q-p}{2}}}\sim v^{\frac{p-q}{2}-+s+1}%
.\label{tau_sc}%
\end{equation}

The proper time is finite $s<\frac{p-q}{2}+1.$ The requirement $s>0$ gives us
$q<p+2$. If, in addition, $q\leq2$, the proper time is finite for any
subcritical particle.

For critical particles one obtains%

\begin{equation}
\tau\sim\int\frac{dv}{v^{\frac{q}{2}}}\sim v^{\frac{2-q}{2}}%
\end{equation}
If $q=2$, we return to the typical situation discussed many times for extremal
black holes starting from \cite{ted}:%
\begin{equation}
\tau\sim\left\vert \ln v\right\vert \text{.}%
\end{equation}

The proper time is finite if $q<2.$

For ultracritical particles%

\[
\tau\sim\int\frac{dv}{v^{c}}\sim v^{1-c}%
\]

The proper time is finite if $c<1.$ As $c>q/2,$ this is possible only if
$q<2.$

Let us stress again that the main property of particles with a finite proper
time is that they cross the horizon. Such particles moving outwardly \ and
participating in frontal collision may appear only if they emerge from a white
hole. However, if the proper time is infinite, then such a particle can move
away from a black (not white) hole horizon. Indeed, as their affine parameter
$\tau$ is infinite, such a trajectory is complete and cannot be extended into
the past, so it could not appear from the region under the horizon. Such
particle can participate in frontal collisions with an ingoing particle. In
particular, this occurs when an outgoing near-critical particle collides with
the usual one. For the Kerr black hole this leads to significant enhancement
of the energy of debris \cite{sch}.

For $c=3/2$ we return to so-called critical particles of class II (see Sec. II
D in \cite{axis}).

\section{Head-on collisions}

We start analysis from consideration of frontal collision and its properties.
Energy of collision is given by%
\begin{equation}
E_{c.m.}^{2}=-(m_{1}u_{1\mu}+m_{2}u_{2\mu})(m_{1}u_{1}^{\mu}+m_{2}u_{2}^{\mu
})=m_{1}^{2}+m_{2}^{2}+2m_{1}m_{2}\gamma\text{,}%
\end{equation}
where $\gamma=-u_{1\mu}u^{2\mu}$ is the Lorentz gamma factor of relative
motion. Substituting the expression for the four-velocity (\ref{4_vel}), we
have%
\begin{equation}
\gamma=\frac{\mathcal{X}_{1}\mathcal{X}_{2}+P_{1}P_{2}}{N^{2}}-\frac
{\mathcal{L}_{1}\mathcal{L}_{2}}{g_{\varphi\varphi}}. \label{gamma}%
\end{equation}

Here we assume that particle 1 is outgoing, while particle 2 is ingoing, so in
(\ref{4_vel}) $\sigma_{1}=-\sigma_{2}=+1$.

The second term in (\ref{gamma}) is regular, so we are interested, when the
first one is unbounded. To analyze the behavior of the first term, let us
consider all possible combinations of different types of particles.

\subsubsection{particle 1 is usual}

This entails that $\mathcal{X}_{1}$ $\sim1$, $P_{1}\sim1.$ Then particle 2 may
be of any type. If it is usual, then $\mathcal{X}_{2}\sim1,$ $P_{2}\sim1$ that
according to (\ref{gamma}) gives us%
\begin{equation}
\gamma\sim v^{-p}\text{.}%
\end{equation}

If the second particle is subcritical, then for such a particle $\mathcal{X}%
_{2}\sim P_{2}\sim v^{s_{2}}$ that makes the gamma factor (\ref{gamma})%
\begin{equation}
\gamma\sim v^{s_{2}-p}\text{.}%
\end{equation}

If the second particle is critical, then for such a particle $\mathcal{X}%
_{2}\sim P_{2}\sim v^{p/2}$ and the gamma factor becomes%
\begin{equation}
\gamma\sim v^{-p/2}.
\end{equation}

If the second particle is ultracritical, then $\mathcal{X}_{2}\sim v^{p/2},$
while $P_{2}\sim v^{c+\frac{p-q}{2}}$, where $c>q/2$ \cite{force23}. Now,
$P_{2}$ is of higher order then $\mathcal{X}_{2},$ so among two terms in
$\mathcal{X}_{1}\mathcal{X}_{2}+P_{1}P_{2}$ the first one is the dominant.
Thus we have%
\begin{equation}
\gamma\sim v^{-p/2}.
\end{equation}

\subsubsection{particle 1 is subcritical}

If the first particle is subcritical,
\begin{equation}
\mathcal{X}_{1},P_{1}\sim v^{s_{1}}%
\end{equation}
with $s_{1}<p/2$.

If particle 2 is usual we return to the case already discussed above.

If particle 2 is subcritical, then $\mathcal{X}_{2},P_{2}\sim v^{s_{2}}$. Thus
we obtain:%
\begin{equation}
\gamma\sim v^{s_{1}+s_{2}-p}\text{.}%
\end{equation}

If the second particle is critical, $\mathcal{X}_{2},P_{2}\sim v^{p/2}$ and we
obtain:%
\begin{equation}
\gamma\sim v^{s_{1}-p/2}\text{.}%
\end{equation}

If particle 2 is ultracritical, then $\mathcal{X}_{2}\sim v^{p/2}$ while
$P_{2}\sim v^{c+\frac{p-q}{2}}.$ As in this case $c>q/2$ \cite{force23}$,$
$P_{2}$ is of the higher order than $\mathcal{X}_{2},$ so among two terms in
$\mathcal{X}_{1}\mathcal{X}_{2}+P_{1}P_{2}$ the first one is the dominant.
Thus we have:%
\begin{equation}
\gamma\sim v^{s_{1}-p/2}\text{.}%
\end{equation}

\subsubsection{particle 1 is critical or ultracritical}

If particle 1 is critical,
\begin{equation}
\mathcal{X}_{1},P_{1}\sim v^{p/2}\,. \label{crit}%
\end{equation}

If particle 2 is usual or subcritical, we return to what was discussed above.

If the second particle is critical, then $\mathcal{X}_{2},P_{2}\sim v^{p/2}$
and one gets%
\begin{equation}
\gamma\sim1\text{. }%
\end{equation}

The same holds for the case when second particle is ultracitical.

If particle 1 is ultracritical, one can check that we will obtain the same
expression for the gamma factor. We summarize all the results in a
corresponding Table \ref{tab_1}:

\begin{table}[ptb]%
\begin{tabular}
[c]{|c|c|c|}\hline
1-st particle & 2-nd particle & $d$\\\hline
Usual & Usual & $p$\\\hline
Usual & Subcritical & $p-s_{2}$\\\hline
Usual & Critical or ultracritical & $p/2$\\\hline
Subcritical & Usual & $p-s_{1}$\\\hline
Subcritical & Subcritical & $p-s_{1}-s_{2}$\\\hline
Subcritical & Critical or ultracritical & $p/2-s_{1}$\\\hline
Critical or ultracritical & Usual & $p/2$\\\hline
Critical or ultracritical & Subcritical & $p/2-s_{2}$\\\hline
Critical or ultracritical & Critical or ultracritical & $0$\\\hline
\end{tabular}
\caption{ Table showing behaviour of gamma factor for frontal collision of
particles of different types. Here $d$ is defined to be degree of gamma
factor's divergence: $\gamma\sim v^{-d}$}%
\label{tab_1}%
\end{table}

\section{Results for different types of horizons}

In this section we are going to summarize corresponding results for different
types of horizons and indicate when UGEF (unbounded growth of energy in
frontal collisions) is possible. We start with non-extremal horizons for which
$q=p=1.$ Any particle moving in such a space-time has a finite proper time
that follows from Section \ref{sec_pr_time}. According to Table \ref{tab_1},
the UGEF phenomenon is possible for all types of particles, except from the
cases when both particles are critical or ultracritical. All cases in which
this effect is possible are summarized in Table \ref{tab_2}. There, along with
the types of particles participating in the effect, we add information about a
behaviour of the proper time for ougoing particle 1. We also indicate for what
kind of a space-time this phenomenon is possible: for a black hole or a white
hole (we remind a reader that if the proper time is finite for the first
particle, then this effect may take place only in a white hole spacetime,
while if it is infinite, this happens for a black hole).

\begin{table}[ptb]%
\begin{tabular}
[c]{|c|c|c|c|}\hline
1-st particle & $\tau_{1}$ & 2-nd particle & BH or WH\\\hline
Usual & F & Usual, subcritical, critical or ultracritical & WH\\\hline
Subcritical & F & Usual, subcritical, critical or ultracritical & WH\\\hline
Critical or ultracricial & F & Usual or subcritical & WH\\\hline
\end{tabular}
\caption{ Table showing all cases in which the UGEF phenomenon is possible for
a non-extremal horizon. In this table the 1-st and the 3-rd columns show a
type of particles for which BSW phenomenon is possible, the 2-nd column shows
how a proper time behaves for the first (outgoing) particle (here "F" means
that a proper time is finite, "D" means that it diverges). The last column
shows for which spacetimes this phenomenon is possible (here "BH" stands for a
black hole, "WH" for a white hole).}%
\label{tab_2}%
\end{table}

If the horizon is extremal (that requires $q=2$), the proper time is finite
only for usual and subcritical particles (moreover, it is finite for all $s$,
as one can see from (\ref{tau_sc})). The UGEF phenomenon, as before, is
possible for all types of particles (except from the case when both particles
are critical, see Table \ref{tab_1}). All cases in which this effect is
possible are summarized in Table \ref{tab_3}.

\begin{table}[ptb]%
\begin{tabular}
[c]{|c|c|c|c|}\hline
1-st particle & $\tau_{1}$ & 2-nd particle & BH or WH\\\hline
Usual & F & Usual, subcritical, critical or ultracritical & WH\\\hline
Subcritical & F & Usual, subcritical, critical or ultracritical & WH\\\hline
Critical or ultracritical & D & Usual or subcritical & BH\\\hline
\end{tabular}
\caption{ Table showing all cases in which the UGEF phenomenon is possible for
an extremal horizon. In this table the 1-st and the 3-rd columns show a type
of particles for which BSW phenomenon is possible, the 2-nd column shows how a
proper time behaves for the first (outgoing) particle (here "F" means that a
proper time is finite, "D" means that it diverges). The last column shows for
which spacetimes this phenomenon is possible (here "BH" stands for a black
hole, "WH" for a white hole).}%
\label{tab_3}%
\end{table}

If the horizon is ultraextremal (by definition, it means that $q>2$), then
only usual (if $2<q<p+2$) and subritical (if $s<\frac{p-q}{2}+1$) particles
have a finite proper time to (or from) the horizon. The UGEF phenomenon, as
before, is possible for all types of particles (except from the case when both
particles are critical, see Table \ref{tab_1}). The results for ultraextremal
horizons are summarized in Table \ref{tab_4}.

\begin{table}[ptb]%
\begin{tabular}
[c]{|c|c|c|c|}\hline
1-st particle & $\tau_{1}$ & 2-nd particle & BH or WH\\\hline
Usual & F if $2<q<p+2$ & Usual, subcritical, critical or ultracritical &
WH\\\hline
Usual & D if $p+2\leq q$ & Usual, subcritical, critical or ultracritical &
BH\\\hline
Subcritical & F if $s<\frac{p-q}{2}+1$ & Usual, subcritical, critical or
ultracritical & WH\\\hline
Subcritical & D if $\frac{p-q}{2}+1\leq s<\frac{p}{2}$ & Usual, subcritical,
critical or ultracritical & BH\\\hline
Critical or ultracritical & D & Usual or subcritical & BH\\\hline
\end{tabular}
\caption{ Table showing all cases in which the UGEF phenomenon is possible for
an ultraextremal horizon. In this table the 1-st and the 3-rd columns show a
type of particles for which the UGEF phenomenon is possible, the 2-nd column
shows how a proper time behaves for the first (outgoing) particle (here "F"
means that a proper time is finite, "D" means that it diverges). The last
column shows for which spacetimes this phenomenon is possible (here "BH"
stands for a black hole, "WH" for a white hole).}%
\label{tab_4}%
\end{table}

It is worth noting that a possible presence of a force is taken into account
in an indirect way - through the number $s$. If the forces are absent, $E$ and
$L$ are conserved, so it follows from (\ref{om_exp}), (\ref{hi}) and
(\ref{L_exp}) that $s=k$. If the forces and corresponding accelerations are
present, $s\neq k$. Then, there is an interplay between the behavior of
acceleration and a type of particle encoded in $s$ \cite{force23}.

We would also like to pay attention to the following subtlety. In any event,
the energy of collision cannot be infinite in a \ literal sense, although it
may be as large as one likes. This is a so-called principle of kinematic
censorship \cite{cens}. It would seem that this principle is violated if, say,
particle 1 crosses the past horizon, emerging from a white hole and collides
with particle 2. However, this would formally mean that collision occurred
exactly on the horizon. Meanwhile, in such a scenario, particle 2 moves
towards the future event horizon, so both particles cross different horizons
and cannot meet, so collision is absent at all. Such a collision can happen
very nearly to the horizon at some $r=r_{h}(1+\delta)$ with $\delta\ll1$.
Then, $E_{c.m.}\rightarrow\infty$ when $\delta\rightarrow0$ but for any small
nonzero $\delta$ this energy is finite, so the principle under discussion is preserved.

\section{Summary and conclusions}

In this work we analyzed the main properties of the frontal high energy
particle collisions (UGEF). To this end, we had to consider main kinematic
properties of colliding particles - such as near-horizon behavior of the
four-velocity and behavior of the proper time. These results allowed us to
analyze behavior of the energy $E_{c.m.}$ in the center of mass frame and to
conclude for which particles this energy grows unbounded and for which it
remains finite. We showed that the UGEF is possible for all cases except from
the case when both particles are critical or ultracritical. However, this does
not fully answer the question concerning a configuration that is needed to
make the phenomenon under discussion possible near black holes. To elucidate
this question, we were led to find, whether the proper time for an outgoing
particle is finite or infinite. If it is finite, this means that such a
particle crossed a horizon before collision. As this particle is outgoing,
this means that such a collision occurs near the horizon of a white hole.
Otherwise, this phenomenon is possible for black holes since an outgoing
particle could be created in the preceding collision (although we did not
delve into the details of a corresponding scenarios). We did this analysis
depending on the type of the horizon (whether it is non-extremal, extremal or
ultraextremal) and found that for non-extremal horizons the UGEF is possible
only for a white hole, for extremal horizons it is possible for a black hole
only if the outgoing particle is critical or ultracritical (while an ingoing
is usual or subcritical), while for ultraextremal horizons this phenomenon is
possible for both black and white holes.

We did not touch upon the question about astrophysical consequences of high
energy particle collisions. Unfortunately, at present first conjectures on
this subject \cite{pir1} - \cite{pir3} are not justified in observations. The
role of white hole elucidated above, even restricts observational perspectives
further. However, at the same time, it extends theoretical significance of
treatment of high energy particle collisions near white holes which for
particular cases were considered earlier \cite{gpwhite}, \cite{white}. They
can also help to understand better why white holes are unstable (that is,
\ however, a separate subject). In any case the processes under discussion are
predicted from the first principles. And, for better understanding, building
the full theory of such phenomena is absolutely necessary for better
understanding and possibility to apply this to the analysis of observational data.

Thus, in combination with the previous results. we managed to classify all
possible scenarios for two mutually complementary kinds of collisions: the BSW
effect \cite{force23} and the UGEF (the present paper), provided particles
move in the equatorial plane. In doing so, we generalized previous
observations made for particular metrics. A corresponding problem for the
ultra-high energy collisions with particles moving on circular orbits will be
considered elsewhere.

\end{document}